\documentclass[10pt]{article}

\usepackage[english]{babel}
\usepackage[latin1]{inputenc}
\usepackage{natbib}
\usepackage{amssymb}
\usepackage{multicol}
\usepackage[fleqn]{amsmath}
\usepackage{epsfig}
\usepackage[normalem]{ulem}
\usepackage{verbatim}
\usepackage{graphicx}
\usepackage{color}
\usepackage{bbm}
\usepackage{dsfont}

\newcommand{\R}{I\!\!R}
\newcommand{\N}{I\!\!N}

\newcommand{\I}{\mathds{1}}

\newcommand{\E}{\mbox{E}}

\newcommand{\1}{\mathbbm{1}}

\newcounter{cptpropo}[part]
\newenvironment{propo}[0]
{\noindent\textsc{Proposition}\,\refstepcounter{cptpropo}\thecptpropo.\it}

\newcounter{cptlemmo}[part]
\newenvironment{lemmo}[0]
{\noindent\textsc{Lemma}\,\refstepcounter{cptlemmo}\thecptlemmo.\it}

\newcounter{cptexo}[part]
\newenvironment{exo}[0]
{\noindent\textsc{Example}\,\refstepcounter{cptexo}\thecptexo.\it}

\title{\Large \bf Elicitation of Weibull priors}
\author{Nicolas Bousquet \\ \vspace{0.5cm} \small EDF Research \& Development \\ \small 6 quai Watier 78401 Chatou - France \\ \small E-mail : nicolas.bousquet@edf.fr}
\date{}

\begin{document}
\maketitle


\paragraph{\it Summary} $-$
{Based on expert opinions, informative prior elicitation for the common Weibull lifetime distribution
usually presents some difficulties since it requires to elicit a
two-dimensional joint prior. We consider here a reliability framework where the
available expert information states directly in terms of prior predictive 
values (lifetimes) and not parameter values, which are less intuitive. The
novelty of our procedure is to weigh the expert information by the
size $m$ of a virtual sample yielding a similar information, the prior being seen as a reference posterior. Thus,
the prior calibration by the Bayesian analyst, who has to moderate
the subjective information with respect to the data information, is
made simple. A main result is the full
tractability of the prior under mild conditions, despite the
conjugation issues encountered with the Weibull distribution. Besides, $m$ is a practical focus point for 
discussion between analysts and experts, and a helpful parameter for leading sensitivity studies and reducing the potential imbalance in posterior selection between Bayesian Weibull models, which can be due to favoring arbitrarily a prior. The calibration of $m$ is discussed and a real example is treated along the paper.   } \\ 


\section{Introduction}

 The versatile Weibull $\cal W(\eta,\beta)$ distribution,  with
density function
\begin{eqnarray*}
f_W(t|\eta,\beta) & = &
\frac{\beta}{\eta}\left(\frac{t}{\eta}\right)^{\beta-1} \exp\left\{
-\left(\frac{t}{\eta}\right)^\beta\right\} \ \I_{\{t\geq 0\}}
\end{eqnarray*}
$(\eta,\beta)\in (0,\infty)^2$, is one of the most popular
distributions in reliability and risk assessment (RRA) and many other fields, mainly chosen for modelling the lifetime $T$ of an industrial
system or component $\Sigma$ \cite{TSI00}. In real-life studies, a Bayesian
framework has often been highlighted when expert knowledge is
available on $\Sigma$ and observed lifetime data ${\bf
t_n}=t_1,\ldots,t_n$ are small-sized and possibly contain missing or
censored values \cite{BACH98}. Such contexts are usually
encountered in industrial studies, 
especially when economical opportunities imply replacing a range of components at the same time, although they could have 
carried on running, and lead to ``polluted" (e.g., censored) lifetime data. In those cases, all relevant sources of knowledge as expert opinion must be taken into account.

Therefore, numerous authors
\cite{SIN86,SIN88,BERG93}  have focused their work on the
elicitation of a joint prior measure $\pi(\eta,\beta)$ that
formalizes the expert knowledge, in order to integrate some
decision-making function over the joint posterior distribution of
these parameters, with density
\begin{eqnarray*}
\pi(\eta,\beta|{\bf t_n}) & = & \frac{{\cal{L}}({\bf
t_n};\eta,\beta) \ \pi(\eta,\beta)}{\displaystyle \iint_{\R^2_+}
{\cal{L}}({\bf t_n};\eta,\beta) \ \pi(\eta,\beta) \ d\eta d\beta},
\end{eqnarray*}
where ${\cal{L}}({\bf t_n};\eta,\beta)$ denotes the data likelihood. There are two main difficulties using the Weibull
distribution. 
First,
 its only conjugate prior distribution is continuous-discrete
 \cite{SOL69} and remains difficult to justify  in real
 problems \cite{KAM05}. Second, the meanings of scale
 parameter $\eta$ and shape parameter $\beta$ greatly differ.
 Their values and correlation remain hard to assess by
 non-statistician
 experts, even though historical results \cite{LAN01} 
 can be used to provide preferential values as a function of the
 behavior of the studied system. The methods proposed by the previous authors can suffer from this second defect and be applicable with difficulty. 
 Therefore Kaminskiy \& Krivtsov \cite{KAM05} recently
provided a simple procedure to elicit a prior $\pi(\eta,\beta)$
using expert knowledge about the mean and standard deviation of the
cumulative distribution function (cdf) $F_W$. They insisted on the fact that these
values are easier to assess than parameter values. \\

However, especially in sensitive areas like nuclear safety \cite{UNW89}, resorting to a Bayesian framework for reliability-based decision-helping implies defending the metholodology of prior elicitation in front of control authorities, the traditional difficulty being the treatment of its subjective aspects \cite{EFR86}. According to most wishes expressed by decisionners in our practice, the elicitation of a ``defendable" prior $\pi(\theta)$  (of any model parametrized by $\theta$ and not only Weibull) should respect the following items: 
(a) the quantity of subjective information can be directly compared, in terms of percentage, to the quantity of objective (data) information; and 
(b) the elicited prior must be unique.  

One could add to this list other wishes as the practical handling of the prior (explicit features and easy sampling), which is of importance for sensitivity studies. Note that the first item requires to give a clear sense to the words ``quantity of information". Statistical definitions like inverse Fisher matrices or Shannon entropies are often too technical to be directly accessible to decisionners. \\

This article addresses those concerns.  In the sequel,  we consider an
alternative elicitation of $\pi(\eta,\beta)$ defined as the reference posterior of virtual data of size $m$ but calibrated from 
lifetime magnitudes directly given by experts, pursuing the worry of 
realism expressed by Kaminskiy \& Krivtsov \cite{KAM05}. Since the virtual size corresponds to an intuitive measure of prior uncertainty, ratios of virtual and observed data sizes bring an understandable sense to the notion of ``relative quantity of information".  It can appear simpler than
standard deviations (or other typical statistical uncertainty
measures) to discuss with non-statistician experts and, especially, lead to more transparent choices to decisionners. \\  

The structure of the paper is as follows. The full prior elicitation is detailed in (the largest) Section \ref{prior.elicitation}. This methodological section focuses on the calibration of hyperparameters, the aggregation of independent expert opinions and the equitability issues between Bayesian Weibull models. Posterior computation is considered in Section \ref{postcomput}.  A  numerical application  on a real case-study is treated along the paper to illustrate the methodology. A Discussion section ends the paper, presenting alternative results and some avenues for future research. \\


\section{Prior elicitation}\label{prior.elicitation}

\subsection{Principle}

The central idea of prior elicitation comes from a simple 
vision of informative expert opinion already suggested by Lindley \cite{LIN83}. We suggest to consider
that a perfect expert opinion should be, roughly speaking, similar
to a real data survey, and  provide an independent and identically distributed (i.i.d.) sample ${\bf
\tilde{t}_m}=(\tilde{t}_1,\ldots,\tilde{t}_m)$ of lifetime data. 
Now, let $\pi^J$ be a well-recognized formal representation of
ignorance (namely, a noninformative prior), perceived as a reference benchmark measure on the parameter space. Assuming ${\bf
\tilde{t}_m}$ is known, the corresponding prior $\pi$ should be the
posterior distribution with density
$\pi^J(\eta,\beta|{\bf \tilde{t}_m})$. \\

\paragraph{Posterior priors.} Priors built as virtual posteriors present some advantages in
subjective Bayesian analysis. First, they are unique since only
defined by $\pi^J$  and the likelihood (often historically chosen
in experiments). Second, the correlation between parameters is
automatically assessed through the Bayes rule. Third, as said before, the ratio
between the numbers of virtual and real data helps to yield an
understandable answer to the (often unclear) question ``what is the
ratio between subjective and objective information'' asked by
cautious decision-makers. Finally, the aggregation of independent expert opinions is simply carried out through successive Bayes  rules, the consensus virtual sample being an aggregation of all virtual data. This avoids choosing opinion pooling rules which can suffer from paradoxes \cite{oha06b}.

Maybe the most famous of such elicited
priors is Zellner's $g-$prior \cite{ZEL86}. 
Among others,
Clarke \cite{CLA96}, Neal \cite{NEA01}, K\'arny et al. \cite{KAR03}, 
Lin et al.
\cite{LIN06} and Morita et al. \cite{MOR07} examined various
quantification of priors using virtual data. Kontkanen et al. \cite{KON98} considered virtual data as practical tools for eliciting priors for Bayesian networks which may require automation in the treatment of their parameters. \\ 

However, because of various factors, especially subjective ones,
the sample ${\bf \tilde{t}_m}$ is not directly elicitable from an
expert, and his or her information on lifetime $T$ must be
summarized through questioning processes \cite{oha06b}.  
This type of elicitation remains simple  for distributions
belonging to the natural exponential family, for which the
resulting { posterior priors} are conjugate (cf. 
\cite{PRE03}, $\S$ 5.3.3), because the virtual sample can be replaced by exhaustive statistics. In
the continuous Weibull case, unfortunately, the only exhaustive statistic
is the full virtual likelihood. Therefore the questioning must be oriented such that it allows the calibration of nonexhaustive virtual statistics. 

\paragraph{Expert questioning.} In a concern of realism, following the ideas promoted by Kadane \& Wolfson \cite{kad98} and especially Percy \cite{PER02} in the field of RRA, we consider that an expert is mainly capable of providing {\it observable} information on $T$, unconditionally to $(\eta,\beta)$. Indeed, the experts are usually not statisticians and should yield
information independently from any parametrization choice (and
even from any sampling model choice) made by a Bayesian analyst \cite{KAR03}. In other terms we assume that any realistic statistical summary of an expert opinion should be defined with respect to its associated prior predictive
density, namely the prior density of plausible lifetime values
\begin{eqnarray}
f_{\pi}(t) & = & \iint_{\R^2_+} f_W(t|\eta,\beta)\pi(\eta,\beta) \ d\eta
d\beta. \label{margdiroo0}
\end{eqnarray}
The density form of de Finetti's representation theorem \cite{def74} is then invoked to ensure the
unicity of priors elicited in this way, under mild conditions of exchangeability for sequences of values $t_1,\ldots,t_n,\ldots$ See 
Press \cite{PRE03}, $\S$ 10.5, for more precisions. 
 Linking typical magnitudes of the observable variable $T$ with statistical specifications can be made through  decision-theoretical arguments. In the following, we consider experts who can answer to a question similar to the following one:
\begin{center}
{\it Can you give estimates of relative costs $(c_1,c_2)$ linked to two reliability-based decisions induced by the two mutually exclusive events $T\leq t_{\alpha}$ and $T>t_{\alpha}$?}
\end{center}
provided $t_{\alpha}$ is given by the analyst (to improve the {\it phenomenon anchoring} of the expert and diminish subjective bias, cf. \cite{TVE74}) and denoting $\alpha=c_1/(c_1+c_2)\in[0,1]$.  Doing so, following the standard criterion of decision theory, namely the {\it expected utility} \cite{ROB01}, the analyst interprets $t_{\alpha}$ as 
 the minimizer of the predictive Bayes risk 
 \begin{eqnarray*}
{t}_{\alpha} & = & \arg\min\limits_{t_0>0} \int_{0}^{\infty} \Lambda(t_0,t|c_1,c_2) f_{\pi}(t) \ dt.
\end{eqnarray*}
defined by some loss function $\Lambda(t_0,t|c_1,c_2)$ between the choice $T=t_0$ and the unknown truth $T=t$, inflicting $c_1$ to the event $t\leq t_0$ (underestimation) and $c_2$ to the contrary event $t> t_0$ (overestimation). The common choice
\begin{eqnarray*}
\Lambda(t_0,t|c_1,c_2) & = & |t-t_0|\left(c_1\cdot 1_{\{t\leq t_0\}} + c_2\cdot 1_{\{t > t_0\}}\right), 
\end{eqnarray*}
which underlies the analyst wants to penalize similarly small and large misestimations \cite{ROB01},  
leads to $t_{\alpha}$ taking the sense of the $\alpha-$order prior predictive percentile 
\begin{eqnarray}
P_{\pi}(T<{t}_{\alpha}) & = & \int_{0}^{{t}_{\alpha}} f_{\pi}(t) \ dt \ = \ \alpha. 
\label{prior.percentile}
\end{eqnarray}
Therefore an alternative equivalent query is, perhaps simpler, {\it what is the risk $\alpha$ for $\Sigma$ to break down before $t_{\alpha}$?} \\ 


In the following, we assume finally that for each available expert, a unique specified couple $(t_{\alpha},\alpha)$ among all elicitable  
can be considered as his or her {\it most trustworthy specification} (MTS) and must be exactly respected in the effective predictive prior modelling. Various reasons can be invoked for this. First, one cannot hope to elicit a prior $\pi(\eta,\beta)$ such that an arbitrary large number of specified couples $(t_{\alpha},\alpha)$ be exactly respected together, because of the limited flexibility of parametric distributions (Berger 1985 \cite{BER85}, chap. 3). Second, a MTS often appears as a reality since experts have usually more difficulty to speak in terms of extreme values rather than values close to the median behavior \cite{oha06b}. Typically, they can share a similar MTS while their extremes can differ (see Example \ref{example.0}).  Other arguments can be related to decision-making: a cautious, {\it conservative} couple $(t_{\alpha},\alpha)$ can be favored by the analyst since the posterior analysis is focused on percentiles of higher, more critical  orders. \\

\begin{exo}\label{example.0}
Table \ref{expertises001}, already used in \cite{BOU06}, summarizes two
prior opinions about the lifetime $T$ (in months) of a device
 belonging to the secondary water circuit of French
nuclear plants. According to a large
consensus in the RRA field, $T$ is assumed to be well described by a Weibull
distribution. Giving a normative sense to extreme events ($90\%$ credibility), these experts were not questionned at the same level of precision.
${\cal{E}}_1$ is a nuclear operator and spoke about a particular
component, in terms of replacement costs. Conversely, ${\cal{E}}_2$ is  a component vendor 
whose opinion took into account a variety of operating conditions. Costs invoked here were mainly related to mass production.  Therefore the two experts can be considered independent. 
Hence the common median appears as a robust specification and is chosen as the MTS for both.  
\end{exo}

\begin{table}[bhtp]
\centering
 \begin{tabular}{lcc}
             & \small Credibility intervals (5\%,95\%) & \small Median value \\
  \hline
\vspace{-0.25cm}
& &\\
\small expert ${\cal{E}}_1$           & \small [200,300] & \small 250 \\
\small expert ${\cal{E}}_2$  & \small [100,500] & \small 250  \\
\hline
\end{tabular}
\caption{Expert opinions about the lifetime $T$ of a nuclear
device (in months).} \label{expertises001}
\end{table}

\subsection{A comfortable prior form}\label{ideal}

Let us choose $\pi^J$ as the Jeffreys prior for Weibull. In more general Bayesian settings, Sun \cite{SUN97} proposed to favor the Berger-Bernardo reference prior \cite{BERG92} since it has
slightly better properties of frequentist posterior coverage.  
But
this prior requires at least $m\geq 2$ to get proper posteriors.
This would be limiting in practice, when $m$ is chosen small
as it could be expected in cautious subjective assessments. Moreover, 
expert knowledge exerts here itself on $T$ and not on any Weibull parametrization, therefore 
it seems relevant that a benchmark $\pi^J$ be parametrization-invariant. 
See
 \cite{CLA96} for a straightforward defence of Jeffreys'
prior in related problems, where a subjective posterior has to be
compared in information-theoretic terms to an objective 
posterior. Thus we consider
\begin{eqnarray*}
\pi^J(\eta,\beta) & \propto & \eta^{-1} \1_{\{\eta\geq 0\}}\1_{\{\beta\geq \beta_0\}}
\end{eqnarray*}
where $\beta_0\geq 0$ is assumed to be fixed by objective reasons, like physical constraints. For instance, a reliability study focusing on industrial components submitted to aging leads to choose $\beta_0=1$ as explained in Bacha (1998) \cite{BACH98}, since involving a time-increasing failure rate. Without particular constraint, $\beta_0=0$.  Scale invariance imposes no other lower bound than 0 for $\eta$. \\

 Denote
${\cal{GIG}}(a,b,\gamma)$ the generalized inverse gamma
distribution with density
\begin{eqnarray*}
f(x) & = & \frac{b^{a} \gamma}{\Gamma(a)} \frac{1}{x^{a \gamma +
  1}}
\exp\left(- \frac{b}{x^{\gamma}}\right) \1_{\{x\geq 0\}}.
\label{GIG}
\end{eqnarray*}
Reparametrizing $x$ in $\mu=x^{-\gamma}$, $\mu\sim{\cal{G}}(a,b)$. 
Then our ideal prior is $\pi(\eta,\beta) =
\pi(\eta|\beta)\pi(\beta)$, such that
\begin{eqnarray}
\eta|\beta & \sim & {\cal{GIG}}\left(m,b({\bf
\tilde{t}_m},\beta),\beta\right),  \label{ideal.eta} \\
\pi(\beta) & \propto & \frac{\beta^{m-1} \ }{{b^{m}({\bf
\tilde{t}_m},\beta)}}  \ \exp\left(m\frac{\beta}{\beta({\bf
\tilde{t}_m})}\right)\1_{\{\beta\geq \beta_0\}}  \label{ideal.beta}
\end{eqnarray}
with ${b({\bf \tilde{t}_m},\beta)  = \sum_{i=1}^m
\tilde{t}^{\beta}_i}$, and $\beta({\bf
\tilde{t}_m})=m(\sum_{i=1}^m \log \tilde{t}_i)^{-1}$. Both
distributions are proper for all $m>0$. The unknown virtual unsufficient statistics 
 $b({\bf \tilde{t}_m},\beta)$
and $\beta({\bf \tilde{t}_m})$ must be replaced in function of available expert information. 
The linkage between the prior form promoted in
(\ref{ideal.eta}-\ref{ideal.beta}) and a MTS $(t_{\alpha},\alpha)$ elicitable for a given expert  
can be done as explained in the next
proposition (proved in Appendix) and its corollary.  \\

\begin{propo}\label{prop1}
For $\alpha\in]0,1[$ and ${t}_{\alpha}>0$, define the
function $b_{\alpha}: \ \N^*\times \R^*_{+}$ by
\begin{eqnarray}
b_{\alpha}(m,\beta) & = & \left({(1-\alpha)^{-1/m}
-1}\right)^{-1} {t_{\alpha}}^{\beta}.
\label{choice.b} \label{choice.beta}
\end{eqnarray}
Then $b_{\alpha}(m,\beta)$  is the only $\beta-$continuous
function such that, being substituted to $b({\bf \tilde{t}_m},\beta)$ in
(\ref{ideal.eta}), Equation (\ref{prior.percentile}) is verified almost surely.
\\
\end{propo}

An immediate and pleasant consequence of replacing 
deterministic expression $b({\bf \tilde{t}_m},\beta)$ by
$b_{\alpha}(m,\beta)$ is that $\pi(\beta)  \propto
\beta^{m-1} \exp\left(-m \beta \log t_{\alpha} +
m{\beta}/{\beta({\bf \tilde{t}_m})} \right)\1_{\{\eta\geq 0\}}$. We recognize here
the general term of a gamma distribution truncated in $\beta_0$. Finally the resulting
prior is
\begin{eqnarray}
\eta|\beta & \sim &
{\cal{GIG}}\left(m,b_{\alpha}(m,\beta),\beta\right), \label{fine.eta} \\
 \beta & \sim &
{\cal{G}}\left(m,\frac{m}{\widetilde{\beta}(m)}\right) \I_{\{\beta\geq \beta_0\}}
\label{fine.beta}
\end{eqnarray}
where $\widetilde{\beta}(m)  =  \left(\log t_{\alpha}
-\beta^{-1}({\bf \tilde{t}_m})\right)^{-1}$. This result deserves some technical remarks.
\begin{description}
\item[(i)] The joint
prior propriety imposes $\beta^{-1}({\bf \tilde{t}_m})<\log
t_{\alpha}$, namely $\prod_{i=1}^{m} \tilde{t}_i <
t^m_{\alpha}$. 
\item[(ii)] The joint prior (\ref{fine.eta}-\ref{fine.beta}) can remain
proper for all $m$ extended on the half-line $\R^*_{+}$. Thus fuzzy or doubtful experts can be graded using $m\leq 1$. This might be 
valuable if a group of $P$ experts is considered as yielding less
information than $P$ i.i.d. data, for instance because they are suspected of mutual influence.
\item[{\bf (iii)}] The ${\cal{GIG}}(a,b,\gamma)$ distribution was
firstly used by Berger \& Sun \cite{BERG93}, $b$ being assessed
independently of $\beta$. However, this choice was made only
because of the posterior conjugate properties conditionally to
$\beta$ (see {$\S$ \ref{postcomput}}), and no meaning was
given to the hyperparameters. Authors like Tsionas \cite{TSI00} adopted similar approaches.
\end{description}

\subsection{Prior calibration}\label{prior.calibration}

In addition to the MTS needed to define the prior form (\ref{fine.eta}-\ref{fine.beta}), supplementary prior information must be available for the calibration of $(m,\widetilde{\beta}(m))$. In the two following paragraphs we consider some cases commonly encountered in RRA. 

\subsubsection{Calibrating $\widetilde{\beta}(m)$}

In RRA, it can occur that the analyst benefits from {\it qualitative information} on the nature of
aging of $\Sigma$. For instance, assuming $\beta_0=0$, if the expert can
 answer the question  {\it what is the
probability $0<\alpha_{\beta_e}<1$ that $\Sigma$ is submitted to
aging?}, one would have a priori $P(\beta<\beta_e)  =
1-\alpha_{\beta_e}$ with $\beta_e=1$ and consequently
\begin{eqnarray}
\widetilde{\beta}(m) & = &  2m \beta_e/\chi^2_{2m}(1-\alpha_{\beta_e})\label{quantile.beta}
\end{eqnarray}
where  $\chi^2_{m}(q)$ is the $q-$order percentile of the
$\chi^2_{m}$ distribution. Other similar questions can be asked
over accelerated aging (${\beta_e}=2$) and extreme cases
(${\beta_e}=5$) reflecting inconceivable kinetics of aging in
industrial applications \cite{BACH98,LAN01}. Otherwise, databases of typical $\beta$ values (e.g., {\it http://www.barringer1.com/wdbase.htm}) can be used to quantify some alternative features of the gamma prior. \\

However, most frequently (as in Example \ref{example.0}), other {\it quantitative information} is available under the form
 of a single or several {\it credibility intervals}, one of whose bounds is the previously chosen MTS.  
We consider $p\geq 1$ supplementary (non-independent) specifications
$\Omega_p=\{t_{\alpha_i},\alpha_i\}_{i\in\{1,\ldots,p\}}$, sorted by increasing order ($\alpha_i<\alpha_{i+1}$ and $t_{\alpha_i}<t_{\alpha_{i+1}}$). Given $m$, calibrating  $\widetilde{\beta}(m)$  under those predictive constraints can be done by minimizing a distance ${\cal{D}}_m(f^*,f_{\pi})$ 
 where 
$f^*$ is a pdf of $T$ respecting exactly the MTS and the specifications listed in $\Omega_p$. To avoid dealing with the infinite number of possible $f^*$, we adopt the approach proposed by Cooke \cite{COO91}:  ${\cal{D}}$ is chosen as the discrete Kullback-Leibler loss function
between required and elicited marginal features
\begin{eqnarray}
\nonumber  {\cal{D}}_m(f^*,f_{\pi}) & = & \sum\limits_{i=0}^{p} P_{f^*}\left(T\in[t_{\alpha_{i}},t_{\alpha_{i+1}}]\right) \log \frac{P_{f^*}\left(T\in[t_{\alpha_{i}},t_{\alpha_{i+1}}]\right)}{P_{f_{\pi}}\left(T\in[t_{\alpha_{i}},t_{\alpha_{i+1}}]\right)} \\
 & = &  \sum\limits_{i=0}^{p}
(\alpha_{i+1}-\alpha_i) \log
\frac{(\alpha_{i+1}-\alpha_i)}{\left(\alpha^{(e)}_{i+1}-\alpha^{(e)}_i\right)}
\label{optimization} 
\end{eqnarray}
where $t_{\alpha_0}=0$, $t_{\alpha_{p+1}}=\infty$, $\alpha_0=\alpha^{(e)}_0=0$,
$\alpha_{p+1}=\alpha^{(e)}_{p+1}=1$, and for $i\in\{1,\ldots,p\}$ 
\begin{eqnarray*}
\alpha^{(e)}_i  & =  & {\displaystyle \iint
F_W\left(t_{\alpha_i}|\eta,\beta\right)  \pi(\eta,\beta) \
d\eta d\beta}.
\end{eqnarray*} 
The convexity of this loss function in its argument $\pi$ and, given $m$ and $\beta$, the one-to-one continuous correspondence between $\pi(\beta|\widetilde{\beta}(m))$ and $\widetilde{\beta}(m)$   allows for a unique solution of the calibration problem
\begin{eqnarray*}
\widetilde{\beta}^*(m) & = & \arg \hspace{-0.25cm} \min\limits_{\pi(.|\widetilde{\beta}(m))} {\cal{D}}_m(f^*,f_{\pi}).
\end{eqnarray*}
From (\ref{quantile.beta}), estimating $\widetilde{\beta}^*(m)$ is similar to select $\alpha_{\beta_e}=0.5$ and  minimizing (\ref{optimization}) in the prior median $\beta^*_e$.  This provides a direct view of the underlying aging and  numerical estimations were found slightly more robust than those of the prior mean, or those of the best order $\alpha^*_{\beta_e}$ if, conversely, $\beta_e$ is fixed. Therefore we temporarily note $\widetilde{\beta}(m)=\widetilde{\beta}_m(\beta_e)$. For a given $m$, a combination of golden section search and successive parabolic interpolation \cite{BRE73} can achieve a robust optimization of $\beta_e$, provided the $\alpha^{(e)}_i$ are smoothly computed at each step of the algorithm. A smooth Monte Carlo estimation can be obtained using a unique importance sampling run $\beta_1,\ldots,\beta_M\sim {\cal{G}}(m,m/\beta_{e,0})$, with large $M$, where $\beta_{e,0}$ is a chosen starting point:  \\
\begin{eqnarray*}
\alpha^{(e)}_i\left(m,\beta_e\right) & \simeq & {\displaystyle 1-\left(\frac{\beta_{e,0}}{\widetilde{\beta}_m(\beta_e)}\right)^m\frac{1}{M}\sum\limits_{j=1}^M \left[{\left(1 + \frac{t^{\beta_j}_{\alpha_i}}{b_{\alpha}(m,\beta_j)}\right)} {\exp\left(\beta_j\left[\frac{1}{\widetilde{\beta}_m(\beta_e)} - \frac{1}{\beta_{e,0}}\right]\right)}\right]^{-m} }.
\end{eqnarray*}
Note that 
\begin{eqnarray*}
\mbox{Err}(m) = {\cal{D}}_m\left(f^*,f_{\pi\left(.|\widetilde{\beta}^*(m)\right)}\right) 
\end{eqnarray*}
 measures the expert incoherency with respect to the predictive Weibull distribution, given a virtual sample of size $m$ in agreement with the expert opinion.
 If $\mbox{Err}(m)$ remains large for many $m$, the Weibull choice for the virtual data (and therefore for any real dataset, provided the expert is relevant for the problem) is at least debatable, not to say probably inappropriate. \\ 




\subsubsection{Calibrating $m$}

The calibration of $m$ must be adapted to the experimental context.
A decisionner can impose a given virtual size  to improve the clarity of the posterior result. For instance,
Marin and Robert \cite{MAR10} proposed to give to the virtual size parameter of Zellner's $g-$prior (on regressors of a gaussian linear regression problem) the value $m=1$ by default. In a similar context, another possibility, proposed by Celeux et al. \cite{CEL06} and Liang et al. \cite{LIA08} among others, is to establish an upper hierarchical level in the Bayesian model by considering $m$ as a random variable for which a weakly informative prior must be elicited. A last possibility is to use $m$ as a discussion tool between the analyst and the expert, since the meaning of $m$ is understandable outside the statistical field. Some heuristic methods in this sense are discussed in \cite{BOU06}. \\

However our aim as a  Bayesian analyst is mainly to measure the strenght of the expert opinion (assumed being correctly reflected through the prior modelling) through $m$. Besides, when the experts are no longer questionable and only a summary of their past opinions remains available, 
 it seems somewhat difficult to elicit a hyperprior on this parameter. Then we suggest that $m$ should be integrated as the minimizer of the expert incoherency risk, namely
\begin{eqnarray*}
m^* & = & \arg\min_{m\geq 0} \mbox{Err}(m).
\end{eqnarray*} 
It is the analyst's decision to minimize this risk on $\N^*$ (not to loose the virtual size meaning) or $\R^*_+$. In our experiments we chose $\R^*_+$ to get the closest prior to the expert opinion. Obviously, one can avoid eliciting a too informative prior by limiting the minimization domain to $(0,n]$.  \\

\begin{figure}
\centering
   \begin{minipage}[c]{.46\linewidth}
    \includegraphics[width=7cm,height=7.5cm]{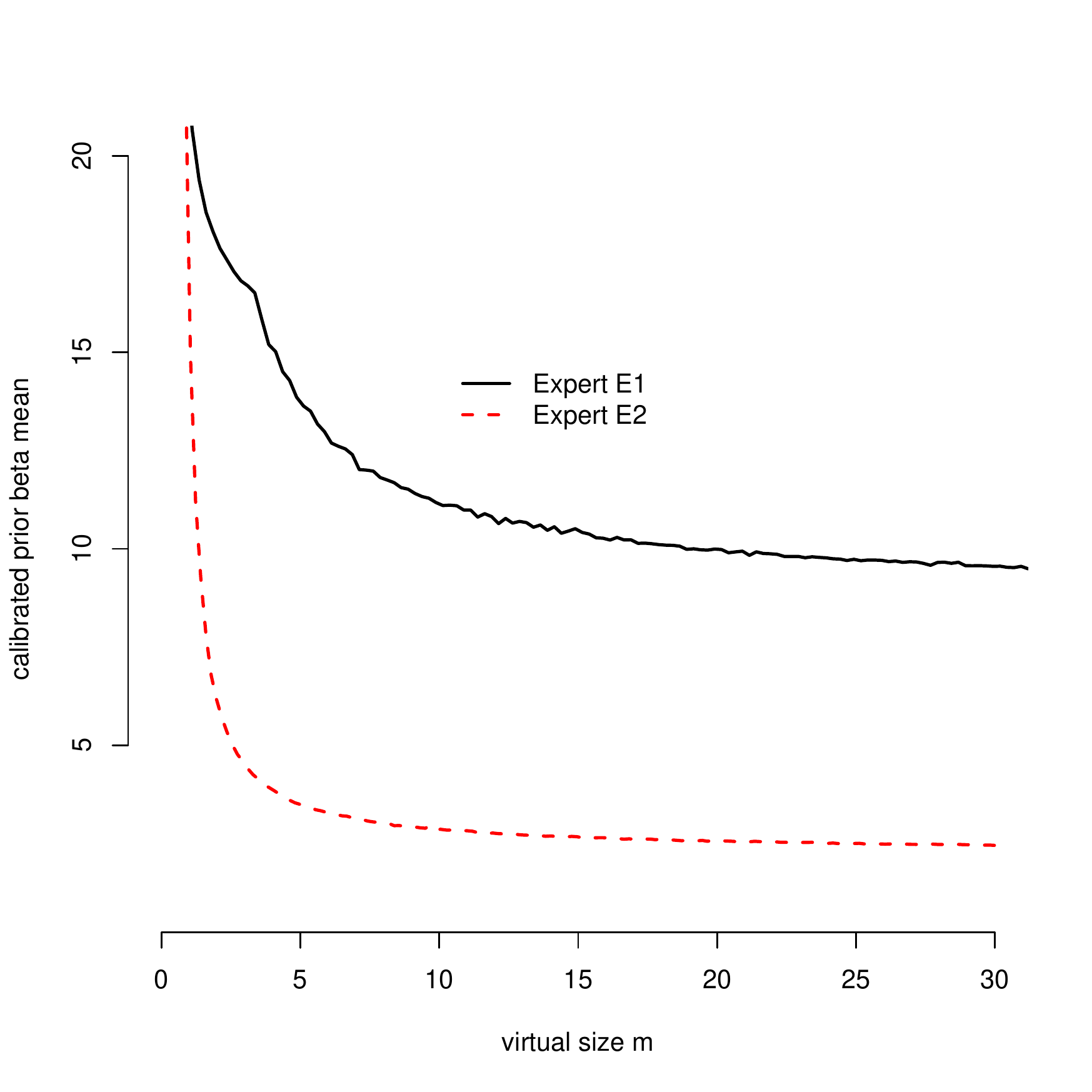}
\caption{Calibrated prior mean $\tilde{\beta}^*(m)$ in function of $m$, for each expert opinion summarized in Table \ref{expertises001}.}
\label{betam}
  \end{minipage} \hfill 
   \begin{minipage}[c]{.46\linewidth}
      \includegraphics[width=7cm,height=7.5cm]{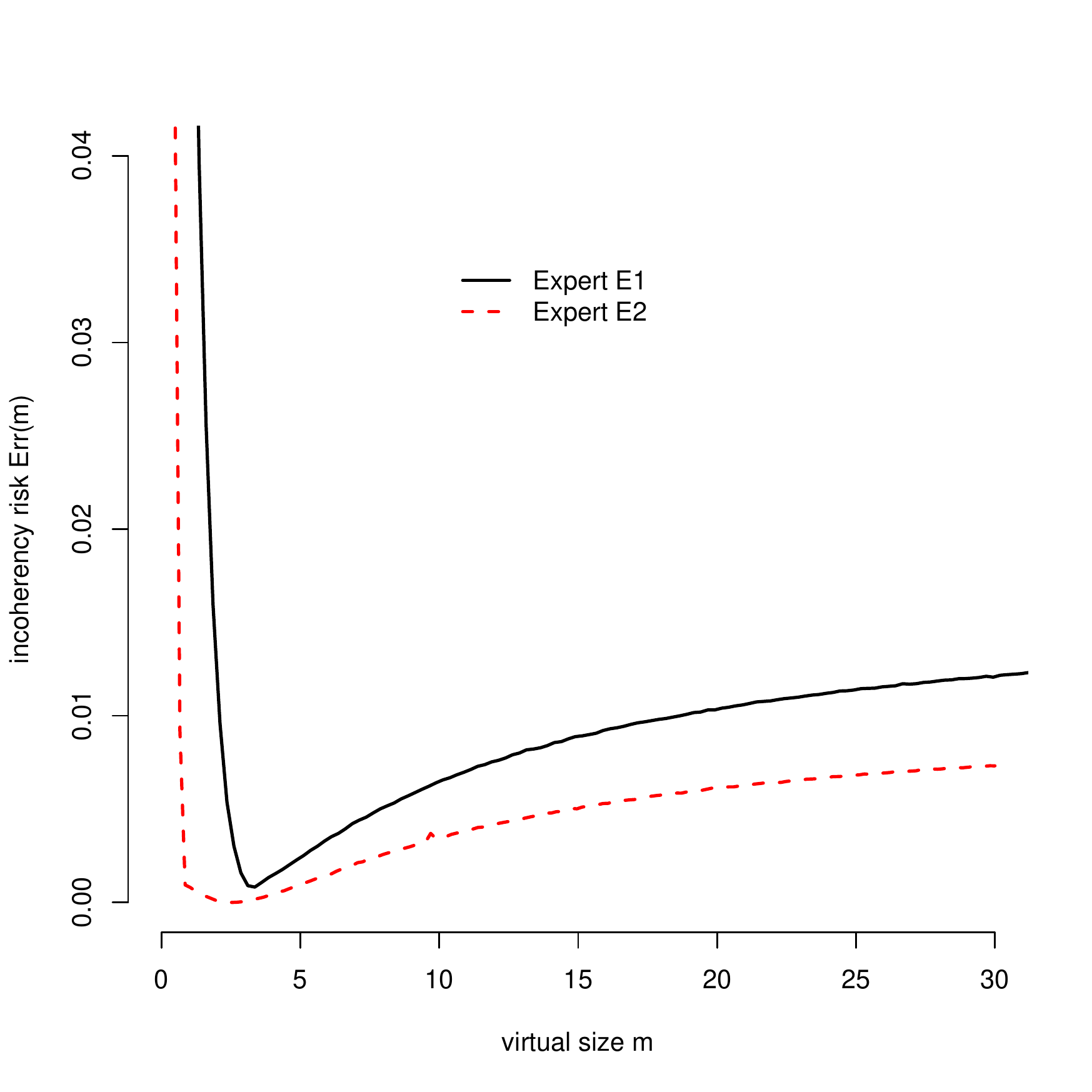}
\caption{Expert incoherency risk $\mbox{Err}(m)$ in function of $m$, for each expert opinion summarized in Table \ref{expertises001}. Minima of plots indicate calibrated values $m^*$.} \label{errm}
  \end{minipage}
    \begin{minipage}[c]{.46\linewidth}
      \includegraphics[width=7cm,height=7.5cm]{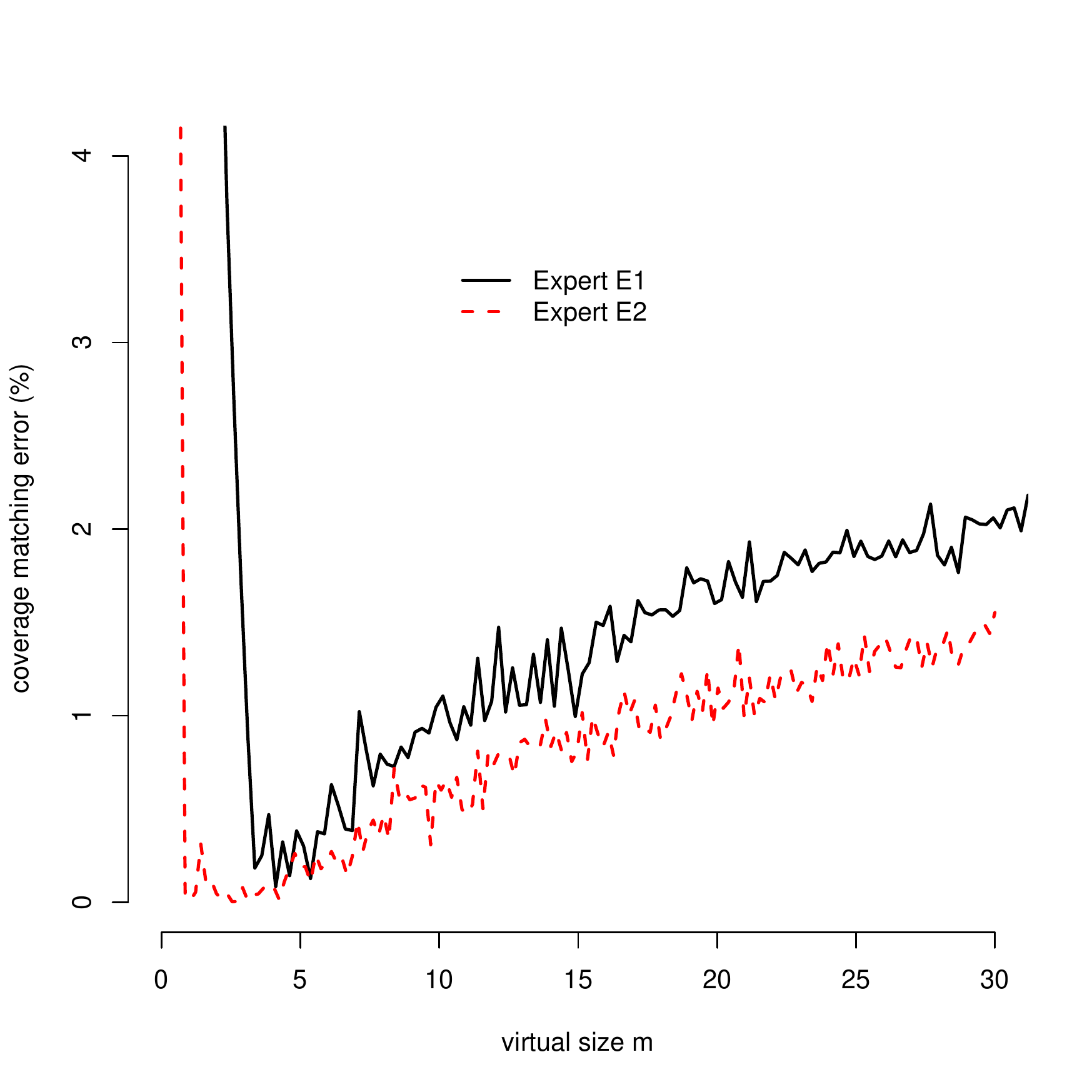}
\caption{Prior predictive coverage matching error in percentage.} \label{cov} 
  \end{minipage} \hfill 
   \begin{minipage}[c]{.46\linewidth}
   \vspace{0.5cm}
      \includegraphics[width=7cm,height=7.5cm]{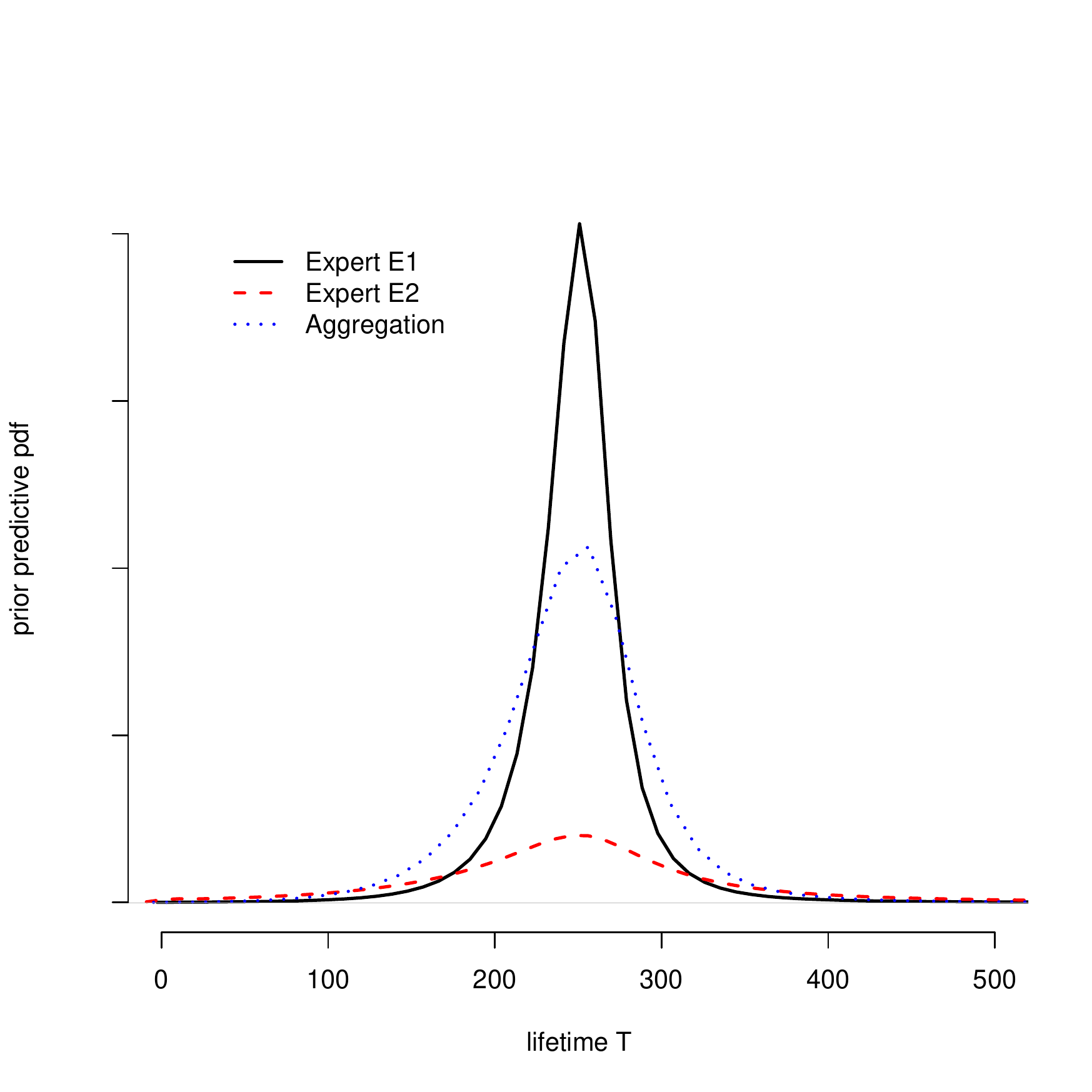}
\caption{Prior predictive densities for both expert opinions and the aggregation of opinions.}\label{aggrego}
  \end{minipage}
\end{figure}

\begin{exo}\label{example.2}(\textsc{\footnotesize{pursuing Example \ref{example.0}}}). 
 For a continuum of values of $m$, we display on Figures \ref{betam} and \ref{errm} the optimized $\tilde{\beta}^*(m)$ and the corresponding risk $\mbox{Err}(m)$, respectively, for both experts. In both cases, the shape of $m\mapsto \mbox{Err}(m)$ allows for a unique solution $m^*$. We find $m^*=3.36$ for expert ${\cal{E}}_1$ and $m^*=2.50$ for expert ${\cal{E}}_2$. This is logical since ${\cal{E}}_1$ is more informative than ${\cal{E}}_2$. 
  However, all values $\tilde{\beta}^*(m)$ for expert ${\cal{E}}_1$ appear unrealistic in an industrial physical context, testifying from exponential uncontrolled aging, assuming the Weibull model is correct. Especially, the calibrated $\tilde{\beta}^*(m^*)=16.5$, which induces a peaked normal behavior of the prior predictive distribution (cf. \cite{DOD06}). On the contrary,  the opinion of expert ${\cal{E}}_2$ remains physically plausible ($\tilde{\beta}^*(m^*)=4.9$) although the underlying aging is still strong.  The corresponding coverage matching error 
$|1 - ({\alpha^{(e)}_{t_1} - \alpha^{(e)}_{t_2}})/{90\%}|$, 
 where $\alpha^{(e)}_{t_i}$ is the effective percentile order for $t_i\in\{100,200\}$ or $t_i\in\{300,500\}$, 
is plotted in Figure \ref{cov} and shows a good adequacy between the wanted and effective credible domains: less than 5\% error in all cases, and less than 0.2\% and 0.004\% when choosing the calibrated $m^*$, for experts ${\cal{E}}_1$ and ${\cal{E}}_2$ respectively. \\

\end{exo}

\subsection{Aggregation of independent expert opinions}\label{aggregation}

In cases when the aggregation of  $i=1,\ldots,p$ priors is chosen as a way to avoid interacting biases in a group  of experts, it yields a similar information to that carried by a global virtual sample, which is the union
of all experts' samples $\bf \tilde{t}_{m_i}$. 
Because they are not explicitly known, one may use
a concatenation of known samples ${\bf
\tilde{s}_{{m_1}}},\ldots,{\bf \tilde{s}_{{m_p}}}$ from another model
${\cal{M}}(\eta,\beta)$ such that their parametric
likelihood $(\eta,\beta)\mapsto \ell({\bf
\tilde{s}_{m_1}},\ldots|\eta,\beta)$ leads to the same inference
 as the whole virtual sample. Indeed, we can show easily (cf. Lemma
\ref{lemma1} in Appendix) that
\begin{eqnarray*}
\pi^J\left(\eta,\beta|{\bf \tilde{t}_{m_1}},\ldots,{\bf
\tilde{t}_{m_p}}\right) & = & \pi^J_{{\cal{M}}}\left(\eta,\beta|{\bf
\tilde{s}_{m_1}},\ldots,{\bf \tilde{s}_{m_p}}\right) \ \propto \
\pi^J(\eta,\beta) \prod\limits_{i=1}^p \ell({\bf
\tilde{s}_{m_i}}|\eta,\beta).
\end{eqnarray*}
Next proposition, proved in Appendix, gives an example of
such a likelihood (said {\it virtual} likelihood) for a single expert opinion. \\

\begin{propo}\label{prop2}
 Consider ${\bf
\tilde{s}_m}=(k_{\alpha,m},\beta_{t_{\alpha},m})$, where
$k_{\alpha,m}  =  ({(1-\alpha)^{-1/m}
-1})^{-1}$ and 
$\beta_{t_{\alpha},m}  = 
{\widetilde{\beta}(m)}/{(1+\widetilde{\beta}(m)\log
t_{\alpha})}$,
as a sample whose components follow independently the
${\cal{G}}(m,m(t_{\alpha}/\eta)^\beta)$ and
${\cal{IG}}(m,m\beta)$ distributions, respectively. Then it 
defines a virtual likelihood for an expert opinion summarized by
$(t_{\alpha},\alpha,\widetilde{\beta}(m))$. \\
\end{propo}

After simple algebra, the resulting prior for all expert opinions is of the same
form (\ref{fine.eta}-\ref{fine.beta}), for which (respecting intuition)  $m  = \sum_{i=1}^p m_i$, $b_{\alpha}(m,\beta) =
\sum_{i=1}^p b_{\alpha_i}(m_i,\beta)$ and
\begin{eqnarray*}
\widetilde{\beta}(m) & = & m \left(\sum\limits_{i=1}^p
\frac{m_i}{\widetilde{\beta}_i(m_i)}\right)^{-1}. \\
\end{eqnarray*}

\begin{exo}\label{example.3}(\textsc{\footnotesize{pursuing Example \ref{example.2}}}). 
Denote $(\pi_1,\pi_2)$ the 
priors calibrated in Example \ref{example.2} for each expert.  Denote $\pi_3$ the aggregating prior. Although $\pi_3$ appears less relevant than $\pi_2$ with respect to a Weibull model in a RRA context, we have no supplementary information to weight the strenght of its corresponding virtual sample in $\pi_3$.  Then, by defect, $\pi_3$ is defined by
\begin{eqnarray*}
m & = & 5.86, \ \ \ \ \ 
b_{\alpha}(\beta) \ = \ 7.33\cdot (250)^{\beta}, \ \ \ \ \ 
\widetilde{\beta} \ = \ 8.30.
\end{eqnarray*}
Corresponding prior predictive densities are plotted on Figure \ref{aggrego}. As it could be expected, the aggregation prior realizes a trade-off between the two priors in the sense it favors the common median according to an intermediate peak, due to the addition of spread virtual data (expert ${\cal{E}}_2$) to 
concentrated virtual data (expert ${\cal{E}}_1$).   
\end{exo}

\subsection{Prior equitability among Weibull models}

Weibull models are often used as bricks for more general reliability models, 
like competing risk models \cite{BER06} or mixtures \cite{TSI02}. Especially, a usual challenge in RRA to choose between exponential and Weibull models. Since the exponential is nested into the Weibull model ($\beta=1$), a simple likelihood ratio test can be carried out in the frequentist framework. A Bayes factor is also easy to compute in our framework. Logically, both models share the same prior elicitation method, with the same MTS. 

Then denote $(m_E,m_W)$ the two corresponding virtual sizes, $(\pi_E,\pi_W)$ the associated priors, and assume the more complex prior $\pi_W$ has been calibrated. How  should $m_E$ be calibrated such that none of the prior Bayesian models is arbitrarily favored in absence of real data? The problem of defining such a {\it prior equitability} to reduce bias in  
posterior selection has been considered by many authors (see \cite{CEL06} for a review), who proposed several rules. 
Celeux et al. (2006) \cite{CEL06} gave decisive arguments to calibrate $\pi_E$ such that it minimizes the Kullback-Leibler  divergence  between predictive distributions 
\begin{eqnarray*}
m^*_E & = & \arg\limits_{m_E}\min \mbox{KL}\left(f_{\pi_W},f_{\pi_E}(.|m_E)\right)
\end{eqnarray*} 
where, after simple algebra,
\begin{eqnarray*}
f_{\pi_E}(t|m_E) & = &  m_E\frac{b_{\alpha}^{m_E}(m_E,1)}{(b_{\alpha}(m_E,1)+t)^{m_E+1}}.
\end{eqnarray*}
 A unique Monte Carlo sampling of $f_{\pi_W}$ can be used to
get a smooth description of the KL divergence and its derivative in $m$, so that a coupled Newton-Raphson method can provide a good estimate of $m^*_E$. This strategy can  be carried out on more complex Weibull models, sorting them through their decreasing order of degree of freedom. If all authors have emphasized the difficulty of this task when nested models are nonconjugate with multidimensional parameters, our framework leads to a rather simple optimization. \\

\begin{figure}
\centering
   \begin{minipage}[c]{.46\linewidth}
    \includegraphics[width=7cm,height=7.5cm]{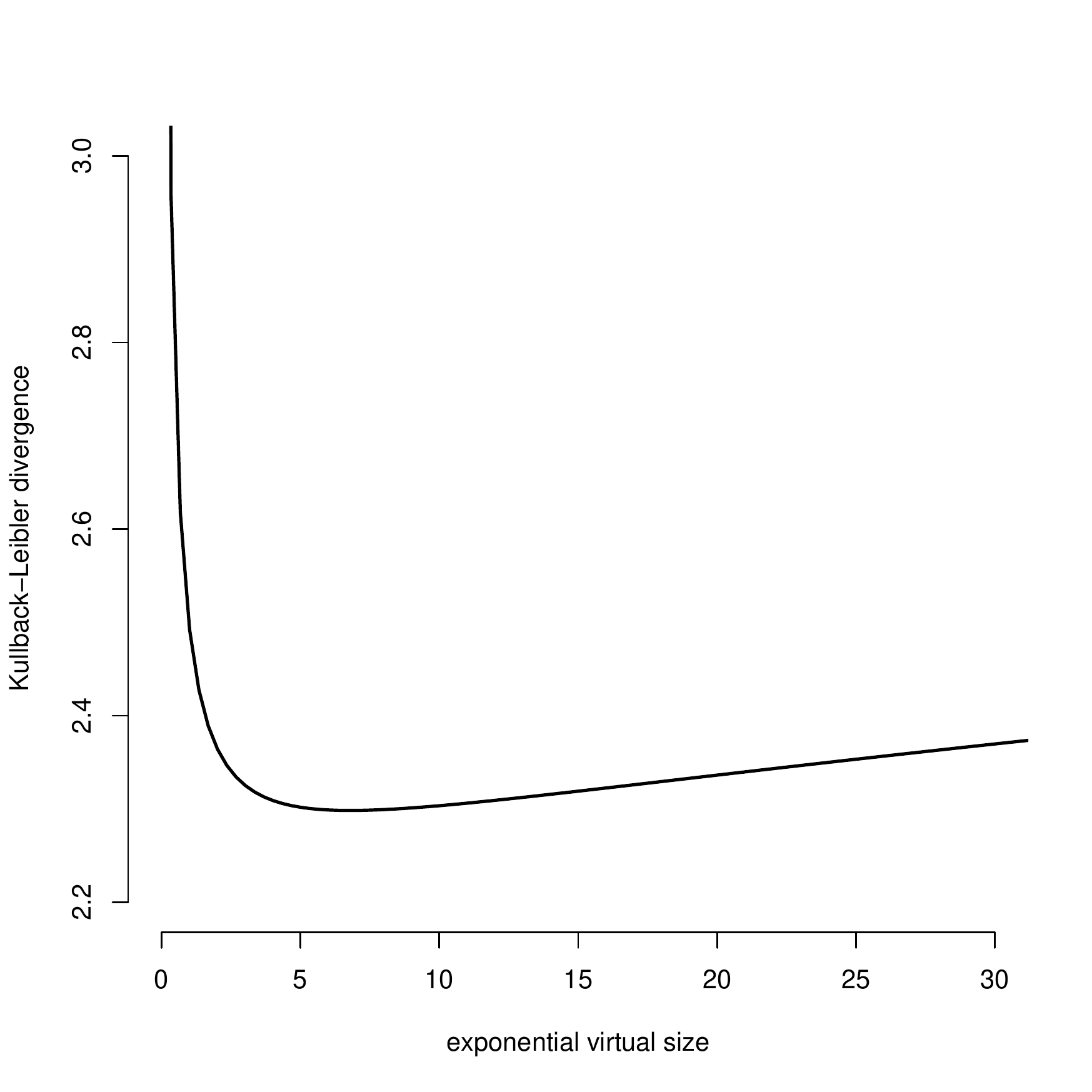}
\caption{KL distance between prior predictive Weibull and exponential distributions dedicated to 
expert ${\cal{E}}_2$'s opinion, in function of the exponential virtual size $m_E$. The minimum is reached in $m^*_E$. }
\label{kl-e2}
  \end{minipage} \hfill 
     \begin{minipage}[c]{.46\linewidth}
    \includegraphics[width=7cm,height=7.5cm]{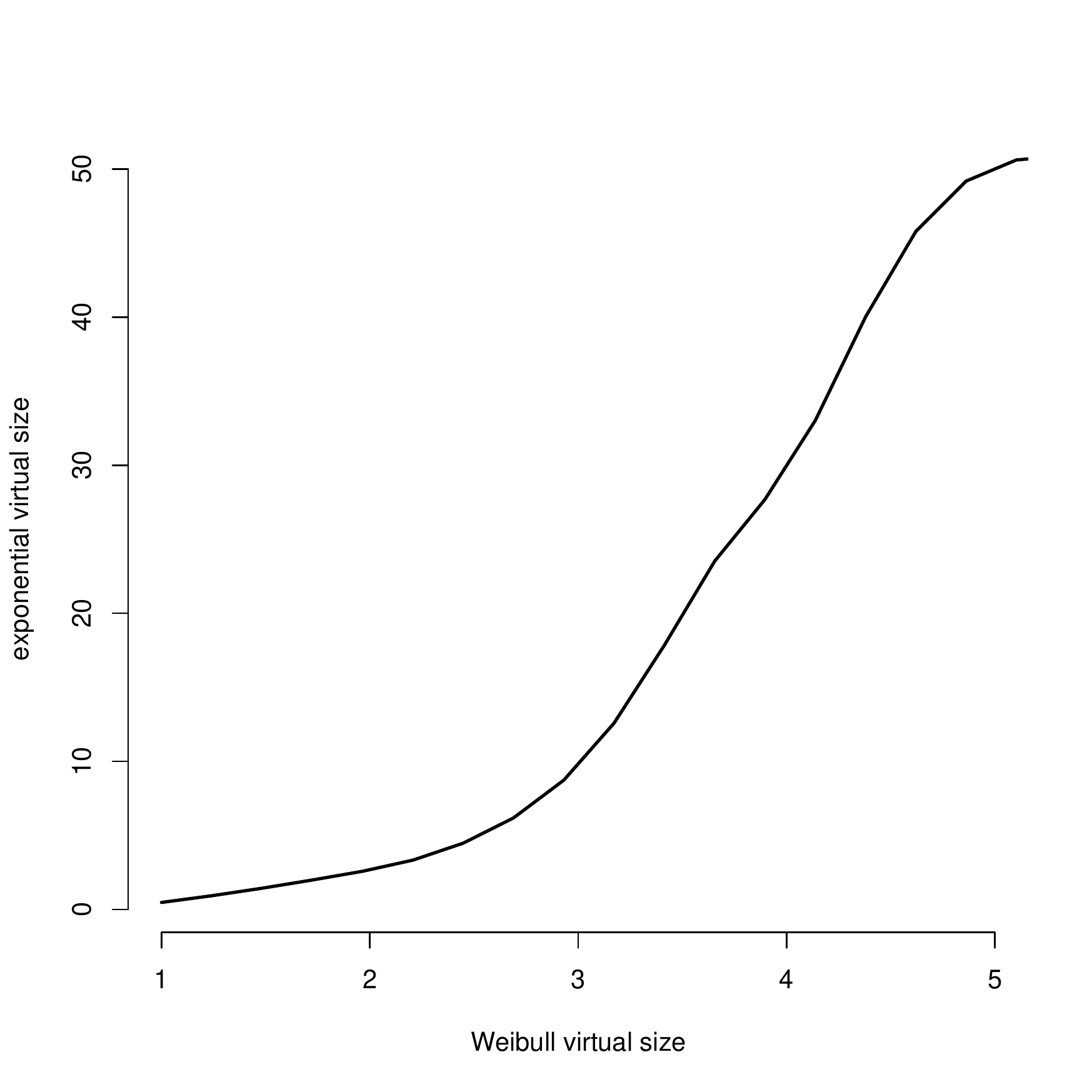}
\caption{Correspondence between Weibull and exponential virtual sizes $(m_W,m^*_E)$ for the expert ${\cal{E}}_2$'s opinion.}
\label{corres0}
  \end{minipage}
\end{figure}

\begin{exo}\label{example.4}(\textsc{\footnotesize{pursuing Example \ref{example.3}}}). 
For various values of $m_E$, the KL divergence is plotted for expert ${\cal{E}}_2$ on Figure \ref{kl-e2}.  The KL 
convexity allows for a unique solution $m^*_E=6.70$. When modifying $m_W$, the correspondence between $m_W$ and $m^*_E$ is plotted in Figure \ref{corres0}.  A similar calculus for expert ${\cal{E}}_1$ leads to a very high value of $m^*_E$ (upper than 200), because any exponential predictive distribution cannot approximate well a peaked normal distribution. An exponential assumption thus appears deeply irrelevant for this expert opinion. 

It could have been expected that $m^*_E\leq m_W$ since the more complex Weibull model should need more data that the exponential one to describe the same prior information. However, the model simplification reduces the global uncertainty in the effective prior predictive distribution. This has a direct impact on the virtual size $m_E$ which varies inversely to uncertainty measures. 
\end{exo}

\section{Posterior inference}\label{postcomput}

Thanks to the considerable development of numerical sampling methods, posterior computation is no longer burdensome in two-dimensional cases. Nonetheless, the conditional conjugation prior properties simplify the work of the Bayesian analyst.  
To be general in the RRA area and in relation with Example \ref{example.5}, we assume that observed data ${\bf
t_n}=(t_1,\ldots,t_n)$ contain $r$ i.i.d uncensored data
$t^{(u)}_1,\ldots,t^{(u)}_r$ and $n-r$
 right-censored data. Denote $
\delta_{\bf t_n}(\beta)  =  \sum_{i=1}^n t^{\beta}_i$ and
$\beta_{\bf t_n}  =  r/\sum_{j=1}^r \log t^{(u)}_i$. Then the joint
posterior distribution has density $\pi(\eta|\beta,{\bf
t_n})\pi(\beta|{\bf t_n})$, such that
\begin{eqnarray*}
\eta|\beta,{\bf t_n} & \sim &  {\cal{GIG}}\left(m+r,b(m,\beta)+\delta_{\bf t_n}(\beta),\beta\right),\\
\pi(\beta|{\bf t_n}) & \propto & \frac{b^m_{\alpha}(m,\beta)\beta^{m+r-1}
}{\left(b_{\alpha}(m,\beta)+\delta_{\bf t_n}(\beta)\right)^{m+r}} \
\exp\left\{-\beta \left(\frac{m}{\tilde{\beta}(m)} -
\frac{r}{\beta_{\bf t_n}}\right)\right\} \ \1_{\{\beta\geq \beta_0\}}.
\end{eqnarray*}
It is enough to obtain approximate posterior sampling of $\beta$ to
get a complete joint sampling (using Gibbs sampling for $\eta$ conditional to $\beta$). This can be made efficiently via the adaptive rejection sampling algorithm from Gilks and Wild \cite{GIL92}. In a noninformative context (i.e., when $m\xrightarrow[]{m>0} 0^+$) which can be easily adapted to a more general setting, Tsionas \cite{TSI00} proposed a gamma instrumental distribution
$\rho(\beta)$ whose mean is calculated to optimize  the acceptance rate. \\

A particular attention must be paid to the existence of posterior moments, especially the first one which defines the Mean Time To Failure (MTTF) in a RRA context. The conditional posterior mean of $\eta$ is 
\begin{eqnarray*}
\E[\eta|\beta,{\bf t_n}] & = & A(r,m,\beta)\left( m\cdot\eta^{\beta}_{e_{|\beta}} + r\cdot\hat{\eta}^{\beta}_{_{|\beta}}\right)^{1/\beta}
\end{eqnarray*}
with  $\eta_{e_{|\beta}}= {k^{1/m}_{\alpha,m}} t_{\alpha}/m$, $\hat{\eta}_{_{|\beta}}  =  r^{-1}\sum_{i=1}^n t^{\beta}_i$ the conditional MLE, and $A(r,m,\beta)  =  {{\Gamma(r+m-1/\beta)}/{\Gamma(r+m)}}$,
so that the MTTF is not defined if $\beta_0\leq 1/(m+r)$. More generally, using an important result of Sun and Speckman \cite{SUN05} (proof of Theorem 5), one can  prove that the $k^\text{th}$ moment of the posterior predictive density
\begin{eqnarray*}
\E\left[T^k|{\bf t_n}\right] & = & \iint_{\R^2} \eta^k \Gamma(1+k/\beta)  \pi(\eta,\beta|{\bf t_n}) \ d\eta d\beta, \\
                             & \propto & \int_{\R_+} \Gamma(1+k/\beta)\Gamma(r+m-k/\beta) \pi(\beta|{\bf t_n}) \ d\beta,
\end{eqnarray*} 
exists only if $\beta_0>k/(r+m)$ for any $k>0$. This result is especially useful in the sense it gives to the analyst a necessary requirement on the prior precision  to justify the practical handling of the posterior predictive distribution through usual statistical summaries, in regards of the information available from really observed data.  In mild conditions ($r\geq 5$ and $k\leq 2$), choosing a defect $\beta_0=k/r$ appears as a practical calculus artifice, and the more justified choice $\beta_0=1$ in aging studies is  sufficient to ensure in practice the existence of posterior predictive moments. \\  
  
\begin{figure}
\centering
   \begin{minipage}[c]{.46\linewidth}
    \includegraphics[width=7cm,height=7.5cm]{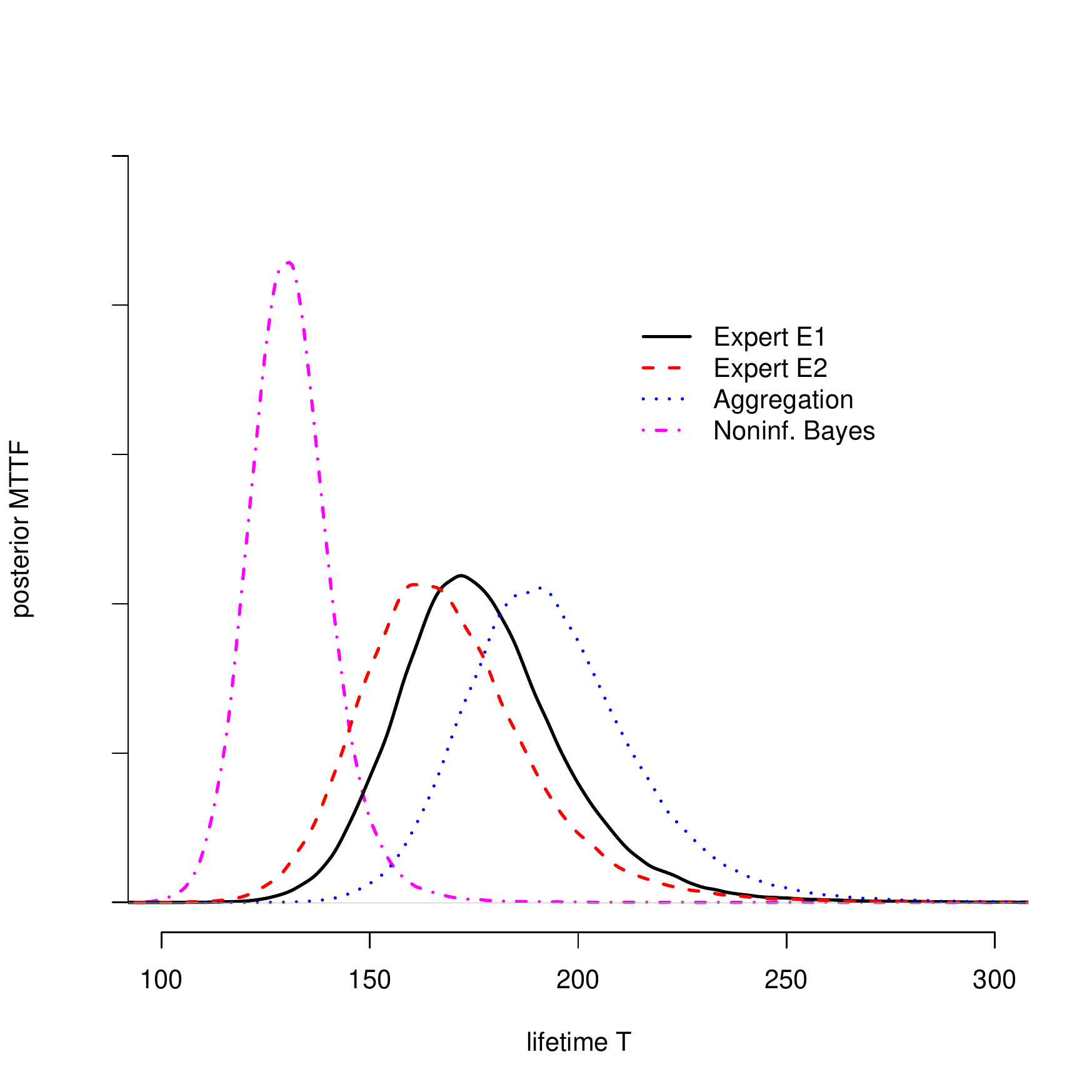}
\caption{Posterior distributions of the MTTF for both experts and their aggregation, and in a noninformative framework. }
\label{mttf1}
  \end{minipage} \hfill 
     \begin{minipage}[c]{.46\linewidth}
    \includegraphics[width=7cm,height=7.5cm]{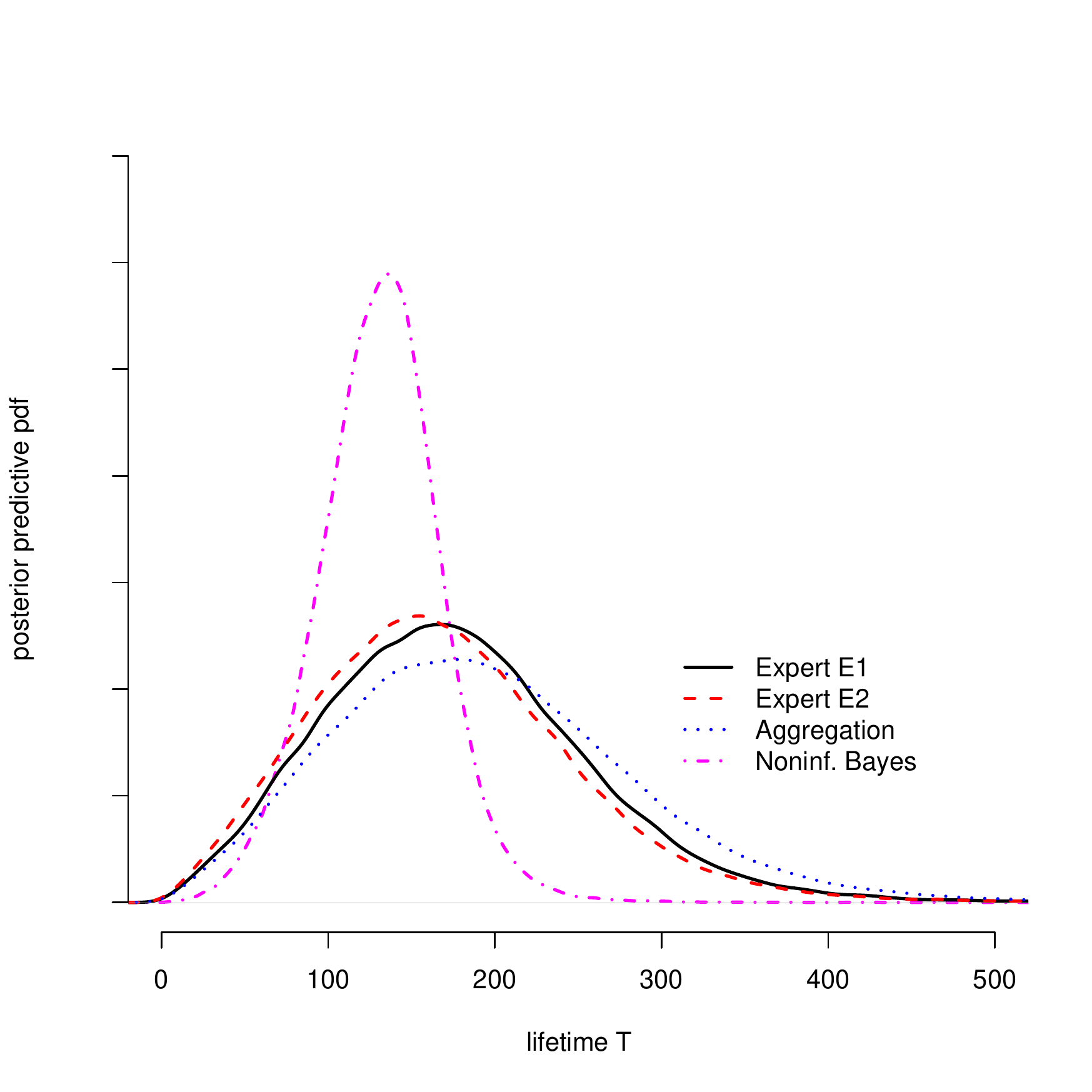}
\caption{Posterior predictive densities for both experts and their aggregation, and in a noninformative framework.}
\label{posterior.predictive.pdf}
  \end{minipage}
\end{figure}

\begin{exo}\label{example.5}(\textsc{\footnotesize{pursuing Example \ref{example.4}}}). 
We consider the right-censored
lifetime data ${\bf t_n}$ ($n=18, r=10$) from Table
\ref{exweib1001bis}. They correspond to
failure or stopping times collected on some similar devices
$\sum$ close to the one considered in Example \ref{example.0}.
 The maximum likelihood estimator (MLE) is
$(\hat{\eta}_n,\hat{\beta}_n)=(140.8,4.51)$ with estimated
standard deviations $\hat{\sigma}_n=(7.3,1.8)$. This strong aging is in agreement with the opinion of expert ${\cal{E}}_2$. 

 Choosing $\beta_0=0.1$ does not modify significantly the calibration of priors for both experts, and the resulting posterior distributions of the MTTF are plotted on Figure \ref{mttf1}. The peak observed when modelling a noninformative expert is due to the concentration of data far from time regions favored by the priors, and logically the aggregating prior, as the most informative, shifts the MTTF to the highest values.  
 
 The calibrated opinions of experts ${\cal{E}}_1$ and ${\cal{E}}_2$ have a relative weight of respectively 34\% and 25\% of the real data information transmitted to the posterior distribution, but, as it could be expected, their  optimism in terms of lifetime has a strong influence on this important function of interest. Note however that due to variance increasing, the left tails of the posterior predictive pdf (Figure \ref{posterior.predictive.pdf}) are upper than the tails of a noninformative posterior predictive pdf. This means that, given a small $t_{\gamma}$ (for instance a replacement time),  the posterior estimation of $\gamma = P(T<t_{\gamma})$ will be slightly overestimated - if we add the experts' opinions - with respect to this given by an only data-driven prediction. On the contrary, when $t_{\gamma}$ is high, the noninformative posterior can appear somewhat too conservative. \\
\end{exo}   

\begin{table}[hbtp]
\centering
\begin{tabular}{ll}
\hline \vspace{-0.35cm}
&\\
\small real failure times & \small 134.9,  \small 152.1,  \small
133.7, \small  114.8,  \small 110.0,
                         \small 129.0,   \small 78.7,   \small 72.8,  \small 132.2,   \small 91.8\\
\small right-censored times  & \small 70.0,  \small 159.5,  \small
98.5,  \small 167.2,  \small 66.8,
                      \small 95.3,  \small 80.9,  \small 83.2   \\
\hline
\end{tabular}
\caption{Lifetimes (months) of nuclear components from secondary
water circuits.} \label{exweib1001bis}
\end{table}

\section{Discussion}\label{discussion}  

The elicitation of a multidimensional prior,
perceived as a reference posterior conditional to virtual data
supposed to reflect a perfect expert opinion, is a practical way of
assessing indirectly the correlations in the parameter space,
coherently with the sampling model. Another important gain is the
possibility of assessing the prior uncertainty in an understandable
way by modulating the virtual size, for instance for sensitivity studies. This indicator of prior information 
might help to increase
the trust of a decision-maker in the posterior beliefs and the
acceptation of Bayesian assessments by control authorities.
Unfortunately, such priors are
often untractable since they require to assess nonexhaustive 
statistics of the virtual data. This is especially the case with the
Weibull models. \\

In this article, however, we showed how this issue can be
overcome, replacing those untractable statistics with functionals
such that the resulting prior answers to the statistical
specifications of the prior knowledge, under the form of
percentiles. Note that other alternatives to percentiles could
have been considered: for instance, following Percy \cite{PER02}, assume an expert can provide
an estimate $t_{e}$ of the marginal MTTF or the mode $\mbox{Md}[T]$, namely
\begin{eqnarray*}
\mbox{MTTF} \ = \ \E[T] & = & \iint_{\R^2_+} \E[T|\eta,\beta] f_W(t|\eta,\beta) \pi(\eta,\beta) \ d\eta d\beta, \\
\mbox{Md}[T] & = & \arg\limits_{t>0}\max \iint_{\R^2_+}  f_W(t|\eta,\beta)
\pi(\eta,\beta) \ d\eta d\beta.
\end{eqnarray*}
The first estimate $t_e$ is thus related to a quadratic loss function $\Lambda(t_0,t|c_1,c_2)$ which assumes the equality of costs $c_1=c_2$ and a penalisation increasing with $|t_0-t|$, while the second can be explained by the limit of a series of binary loss functions \cite{ROB01}. Following the principle described in Proposition \ref{prop1}, one should replace $b({\bf \tilde{t}_m},\beta)$ in
(\ref{ideal.eta}) by, respectively (see \cite{BOU06} for details),
\begin{eqnarray*}
 b(m,\beta) &  = &
\left(\frac{\Gamma(m)}{\Gamma(1+1/\beta)\Gamma(m-1/\beta)}\right)^{\beta}
t^{\beta}_{e}, \\
 b(m,\beta) & = & \left(\frac{m\beta+1}{\beta-1}\right)
t^{\beta}_e.
\end{eqnarray*}
However, 
$\pi(\beta)$ is no longer (but remains close to) a gamma density, and furthermore these two specifications require conditions over $m$ and the domain of variation of $\pi(\beta)$ to be usable. Indeed, one must guarantee $\beta_0>1$ 
to ensure $ b(m,\beta)$ is well defined when $t_e$ is specified as a unique mode. This is coherent with 
the Weibull features, since the Weibull distribution has a unique positive mode if and only if $\beta>1$. As explained before, assuming aging is an equivalent prior constraint placed on the model. 

 Since assuming a prior predictive percentile can be provided by an oriented questioning, and fortunately leads to an
explicit and versatile joint prior on Weibull
parameters, we suggest Bayesian reliability analysts should favor, as much as possible, this kind of elicitation. This agrees with the 
vision historically promoted by Berger \cite{BER85} (chap. 3) and Percy \cite{PER02}, who considered that quantile-based approaches  
pose among best elicitation methods, the estimation  of probabilities of localization in given areas being simpler for experts than the assessment of statistical moments.\\

Along the paper some remaining issues and limitations of the prior modelling have been evoked, which are now discussed as potential avenues for future researches. These researches, besides, 
will be dedicated to extend this methodology to other models which are often used in reliability studies, especially extreme value models whose links with Weibull distributions are well known. \\

It appears firstly that checking for the appropriateness of the Weibull model with respect to the virtual data is a crucial task, since it allows to reveal divergences between an expert opinion and the common sense of reliability practicioners when assuming the Weibull distribution for the lifetime of an industrial component submitted to aging. As the case for expert ${\cal{E}}_1$ illustrates, providing a small prior credibility interval can underlie unrealistic values for the Weibull parameters. This agrees with the well-known behavior of RRA experts of underestimating  their self-uncertainty \cite{LAN01}. Therefore we suggest that spreading the prior credibility orders such that a qualitative requirement is reached (e.g., $P_{\pi}(\beta<2)=0.5$) can give a more reasonnable summary of the real knowledge of $\Sigma$. \\
  
Another issue deals with the remaining uncertainty in expert opinion. In this article, we proposed a simple definition of the ratio of subjective and objective information based on virtual and real data sizes, understandable outside the community of statistics. It is clear however that information quantities updated through Bayesian inference are also strongly dependent on the possible {\it conflicting issues} between prior and real data, in the sense that both can favor regions of the sample and parameter spaces which are far from each other \cite{EVA06,BOU08}. Tools proposed in these two references should be carried out to check the internal coherency of the Bayesian model beforehand. \\ 

Besides, we did not consider here the remaining difficulties occuring when the expert can be suspected of bias, for instance because of {\it motivational reasons} \cite{BEN82} or dependency within a group. Many other tools, based on test experiments, have been proposed in the literature in order to quantify those biases (e.g. \cite{SIN86,SIN88,LAN01}). However, in absence of supplementary information, we prefered respecting the summarized expert opinions the best we could in our case-study. \\

Thus, the case of dependent experts has not been treated in this paper, since it remains controversial in Bayesian statistics \cite{oha06b} and probably deserves specific studies in the RRA area.  
Pursuing our view, two experts are dependent  if they
share  virtual data from their past experience, or if a part of a virtual sample is produced 
dependently from the other sample. Dependency could then be introduced through
a hierarchical mechanism of data production, in the spirit of the {\it supra-Bayesian} approaches promoted by Lindley \cite{LIN83}. This theme will also be considered in our next works.   \\

In future studies, it will be necessary too to provide some calibration 
tools to take into account the expert uncertainty and bias. A first avenue can simply be to add a hierarchical level conditional to $m$. Indeed, we assumed here that the expert subjectivity mainly lies in the self-estimation of the costs associated with a reliability decision, and thus (using notations from $\S$ \ref{prior.calibration}) in the estimation of sorted orders $\alpha_1<\ldots<\alpha_p$. Let us denote $\tilde{\alpha}_1<\ldots<\tilde{\alpha}_p$ these prior estimates. If we pursue the virtual size idea, it appears logical to consider the $\alpha_i$ as correlated random variables such that, a priori,
\begin{eqnarray*}
\alpha_1,\alpha_2-\alpha_1,\ldots,\alpha_{p}-\alpha_{p-1},1-\alpha_p & \sim & {\cal{D}}_{ir}\left(\nu_1,\ldots,\nu_{p+1}\right)
\end{eqnarray*} 
with $\nu_i-1$ being the number of virtual ``past" observations of the event $t_{\alpha_{i-1}}\leq T \leq t_{\alpha_{i}}$, which induces $\sum_{i=1}^{p+1} \nu_i  =  m + p + 1$, the Dirichlet distribution appearing naturally from well-known conjugation properties. Imposing $\E[\alpha_{i}-\alpha_{i-1}]=\tilde{\alpha}_{i}-\tilde{\alpha}_{i-1}$ leads to elicit
\begin{eqnarray*}
\nu_i & = & (m+p+1)\left(\tilde{\alpha}_{i}-\tilde{\alpha}_{i-1}\right).
\end{eqnarray*} 
The obvious correlation between prior estimates $\tilde{\alpha}_i$ threatens to underestimate the prior uncertainty of the $\alpha_i$, so that it appears more appropriate to help the expert providing {\it conditional} probabilities by answering to the following question: {\it what is the risk $\alpha_i-\alpha_{i-1}$ for $\Sigma$ to break down before $t_{\alpha_i}$ knowing $\Sigma$ still runs after $t_{\alpha_{i-1}}$?} Thus, the randomization of values $(\alpha_1,\ldots,\alpha_p)$ imposes indirectly a prior distribution $\pi(\widetilde{\beta}^*(m))$ rather than a single value, and consequently, $m$ should now be calibrated as the minimizer of the {\it expected} expert incoherency risk
\begin{eqnarray*}
m^* & = & \arg\min_{m\geq 0}  \E_{\pi(\widetilde{\beta}^*(m))}\left[ {\cal{D}}_m\left(f^*,f_{\pi\left(.|\widetilde{\beta}^*(m)\right)}\right) \right].
\end{eqnarray*} 
But this calibration remains in facts difficult to carry out, since the untractability of $\pi(\widetilde{\beta}^*(m))$ imposes a double Monte Carlo approximation coupled to an optimization strategy.  Our next research will focus on simplifying this computational work.  \\

Finally, we remind to the reader that other elicitation approaches of percentile orders are possible, mainly based on the establishement of expert preferences on a series of bettings such that $\alpha$ is not directly estimated but is progressively bounded \cite{oha06b}. Such methods are robust with respect to perturbations of the often criticized expected utility criterion, since they can lead to results that are independent of the behavior of the expert face to his or her self-perception of the risk \cite{ABD00}. It thus should be worthy to adapt our modelling and the calibration aspects to this type of elicitation, which could lead to more cautious, credible statistical features of the prior modelling.

\section*{Acknowledgements}

The author thanks  Gilles Celeux (INRIA), Eric Parent and Merlin Keller (ENGREF)  for numerous enriching
discussions, advices and references. He thanks Francois Billy, 
Emmanuel Remy, Alberto Pasanisi (EDF R\&D) too for fruitful discussions about the specific issues raised by
the industrial context of the study. 


\bibliographystyle{plain}

\begin{thebibliography}{}

\small

\bibitem{ABD00} Abdellaoui, M. (2000). Parameter-free elicitation of utilities and probability weighting functions, {\it Management Science}, 46: 1497-1512. 






\bibitem[Bacha et al., 1998]{BACH98}  Bacha, M., Celeux, G., Id\'ee, E., Lannoy, A.
and Vasseur, D. (1998). {\it Estimation de mod\`eles de dur\'ees
de vie fortement censur\'ees}, Eyrolles (in French).


\bibitem[Benson and Nichols (1982)]{BEN82} Benson, P.G., and Nichols, M.L. (1982). An investigation of motivational bias in subjective predictive probability distribution. {\it Decision Sci.}, {\bf 13}: 10-59. 

\bibitem[Berger, J.O.]{BER85} Berger, J.O. (1985). {\it Statistical Decision Theory and Bayesian Analysis}. Springer-Verlag, New York. 


\bibitem[Berger, J.O and Bernardo, J.M. (1992)]{BERG92} Berger, J.O and Bernardo, J.M. (1992). {On the development
of reference priors (with discussion)}. In: J.M. Bernardo, J.O.
Berger, A.P. Dawid and A.F.M. Smith, Eds., {\it Bayesian
Statistics 4}, Oxford University Press: 35-60.

\bibitem[Berger, J.O. and Sun, D. (1993)]{BERG93} Berger, J.O. and Sun, D. (1993). {Bayesian analysis for the
Poly-Weibull Distribution}, {\it J. Amer.
Statis. Assoc.}, {88}: 1412-1418.

\bibitem{BER06} Bertholon, H., Bousquet, N., Celeux, G. (2006). An alternative competing risk model to the Weibull distribution in lifetime data analysis, {\it Lifetime Data Analysis}, 12: 481-504. 


\bibitem[Bousquet, N. (2006)]{BOU06} Bousquet, N. (2006). {A Bayesian analysis of industrial
lifetime data with Weibull distributions}, {Research Report
RR-6025}, INRIA.


\bibitem[Bousquet, N. (2008)]{BOU08} Bousquet, N. (2006).
Diagnostics of prior-data agreement in applied Bayesian analysis.
{\it J. Appl. Statist.}, 35: 1011-1029. 




\bibitem{BRE73} Brent, R. (1973). {\it Algorithms for Minimization without Derivatives}. Englewood Cliffs N.J.: Prentice-Hall.




\bibitem{CEL06} Celeux, G., Marin, J.M., Robert, C.P. (2006). Sélection bay\'esienne de variables en r\'egression lin\'eaire. {\it Journal de la Soci\'et\'e Fran{\c c}aise de Statistique}, 147: 59-79 ({\it in French}). 


\bibitem{CLA96} Clarke, B.S. (1996). {Implications of reference priors for
prior information and for sample size}, {\it J. Amer. Statis. Assoc.}, {91}: 
173-184.



\bibitem[Cooke (1991)]{COO91} Cooke, R.M. (1991). {\it Experts in Uncertainty: Opinion and Subjective Probability  in Science}. New York: Oxford University Press.








\bibitem{def74} De Finetti, B. (1937). La pr\'evision: ses lois logiques, ses sources subjectives. {\it Annales de l'Institut Henri Poincar\'e} (in French), {7}: 1-68.


\bibitem{DOD06} Dodson, B. (2006). {\it The Weibull Analysis Handbook (2nd ed.)}. ASQ Quality Press, p. 7.  

\bibitem{EFR86} Efron, B. (1986). Why isn't everyone a Bayesian? (with Discussion), {\it The American Statistician}, 40: 1-11.




\bibitem[Evans, M., \& Moshonov, H. (2006)]{EVA06} Evans, M., \& Moshonov, H. (2006). Checking for prior-data conflict. {\it Bayesian analysis}, 1: 893-914. 








\bibitem{GIL92} Gilks, W.R., Wild, P. (1992). Adaptive rejection sampling for Gibbs sampling. {\it Applied Statistics}, 41: 337-348. 


\bibitem{kad98} Kadane, J.B., Wolfson, J.A. (1998). {Experiences in
elicitation}, {\it The Statistician}, {47}: 3-19.

\bibitem[Kaminskiy, M.P., and Krivtsov, V.V. (2005)]{KAM05} Kaminskiy, M.P., and Krivtsov, V.V. (2005).
A Simple Procedure for Bayesian Estimation of the Weibull
Distribution, {\it IEEE Trans. Reliability}, 54: 612-616.


\bibitem[K\'arny et al. (2003)]{KAR03} K\'arn\'y, M.,  Nedoma, P., Khailova, N., Pavelkov\'a, L. (2003). Prior information in structure estimation. {\it IEEE Proceedings in Control Theory and Applications}, 150: 643-653. 

\bibitem[Kontkanen et al. (1998)]{KON98} Kontkanen, P., Myllym\"aki, P., Silander, T., Tirri, H., Gr\"unwald, P. (1998). Bayesian and information-theoretic priors for Bayesian networks parameters. {\it Lecture notes in computer science}. {\it In} the proceedings of the ECML-98 European conference on machine learning, Chemnitz, Germany, 1398: 89-94.      



\bibitem[Lannoy, A. and Procaccia, H. (2001)]{LAN01} Lannoy, A. and Procaccia, H. (2001). {\it L'utilisation du jugement d'expert en s\^uret\'e de fonctionnement}, Tec \& Doc (in French).





\bibitem{LIA08} Liang, F., Paulo, R., Molina, G., Clyde, M., Berger, J. (2008). Mixtures of g-priors for Bayesian variable selection. {\it J. American Statist. Assoc.}, 103: 410-423. 




\bibitem[Lin, X., Pittman, J. and Clarke, B. (2007)]{LIN06} Lin, X., Pittman, J. and Clarke, B. (2007). Information Conversion, Effective Samples, and Parameter Size. {\it IEEE Trans. Info. Theory}, 53: 4438-4456.  

\bibitem[Lindley (1983)]{LIN83} Lindley, D.V. (1983). Reconciliation of probability distributions. {\it Operations Research}, 31: 866-880. 





\bibitem{MAR10} Marin, J.M., Robert, C.P. (2010). Les bases de la statistique bay\'esienne. Rapport des Universit\'es Montpellier II \& Dauphine - CREST ({\it in French}).  


\bibitem[Morita et al. (2007)]{MOR07} Morita, S., Thall, P.F. and
Mueller, P. (2007). Determining the effective sample size of a
parametric prior. {\it UT MD Anderson Cancer Center Department of
Biostatistics}, Working Paper Series. Working Paper 36.


\bibitem[Neal (2001)]{NEA01} Neal, R.M. (2001). Transferring prior information between models using imaginary data. Technical Report 0108, Dept. Statistics, Univ. Toronto.





\bibitem{oha06b} O'Hagan, A.,  Buck, C. E., Daneshkhah, A., Eiser, J. R., Garthwaite, P. H.,
Jenkinson, D. J., Oakley, J. E., Rakow, T. (2006), {\it Uncertain
Judgements: Eliciting Expert Probabilities}. John Wiley and Sons,
Chichester.


\bibitem{PER02} Percy, D.F. (2003). Subjective Reliability Analysis Using Predictive Elicitation. In: {\it Mathematical and Statistical Methods in Reliability}, B.H. Lindqvist \& K.A. Doksum (eds).  {\it Quality, Reliability \& Engineering Statistics 7}, World Scientific Publishing Co.: Singapore, pp. 57-72.





\bibitem{PRE03} {Press, S.J.} (2003). {\it Subjective and Objective
Bayesian Statistics} (second edition), New York: Wiley.


\bibitem[Robert, C.P. (2001)]{ROB01} Robert, C.P. (2001). {\it The Bayesian Choice. A
Decision-Theoretic Motivation} (second edition), Springer.




\bibitem[Singpurwalla, N.D. and Song, M.S. (1986)]{SIN86} Singpurwalla, N.D. and Song, M.S. (1986). {An analysis of
Weibull lifetime data incorporating expert opinion}, in {\it
Probability and Bayesian Statistics} (R.Viertl ed.), Plenum
Pub.Corp.: 431-442.

\bibitem[Singpurwalla, N.D. (1988)]{SIN88} Singpurwalla, N.D. (1988). An interactive PC-Based
procedure for reliability assessment incorporating expert opinion
and survival data, {\it J. Amer. Statis. Assoc.}, {83}: 43-51.

\bibitem[Soland, R. (1969)]{SOL69} Soland, R. (1969). {Bayesian analysis of the Weibull
process with unknown scale and shape parameters}, {\it IEEE
Transactions on Reliability}, {18}: 181-184.

\bibitem[Sun (1997)]{SUN97} Sun, D. (1997). {A  note on noninformative priors for
Weibull distributions}, {\it J. Statist. Planning and Inference},
{61}: 319-338.

\bibitem[Sun and Speckman (2005)]{SUN05} Sun, D., Speckman, P.L. (2005). {A  note on the nonexistence of posterior moments}, {\it Can. J. Statist.},
{33}: 591-601.


\bibitem[Tsionas, 2000]{TSI00} Tsionas, E.G. (2000). Posterior analysis, prediction and reliability in three-parameter Weibull distributions, {\it Commun. Statist. - Theory Meth.}, 29(7): 1435-1449.

\bibitem[Tsionas (2002)]{TSI02} Tsionas, E.G. (2002). Bayesian analysis of finite mixtures of distributions, 
{\it Commun. Statist. - Theory Meth.}, 31(1): 37-48.  

\bibitem{TVE74} Tversky, A., Kahneman, D. (1974). Judgment under Uncertainty: Heuristics and Biases, {\it Science}, 185: 1124-1131. 

\bibitem{UNW89} Unwin, S.D., Cazzoli, E.G., Davis, R.E., Khatib-Rahbar, M., Lee, M., Nourbakhsh, H., Park, C.K., Schmidt, E. (1989). An information-theoretic basis for uncertainty analysis: application to the QUASAR severe accident study, {\it Reliability Engineering \& System Safety}, 26: 143-162.





\bibitem[Zellner, A. (1986)]{ZEL86} Zellner, A. (1986). {On assessing Prior Distributions and Bayesian Regression analysis
with g-prior distribution regression using Bayesian variable
selection}, In {\it Bayesian inference and decision techniques :
Essays in Honor of Bruno De Finetti}: 233-243, North-Holland,
Elsevier.



\end{thebibliography}

\section*{Appendix}

\paragraph{Proof of Proposition \ref{prop1}.}
Denote ${\displaystyle f_m(\beta)  = \frac{\beta^{m-1} \
}{{\left(b_{\alpha}(m,\beta)\right)^m}} \
\exp\left(m\frac{\beta}{\beta({\bf \tilde{t}_m})}\right)}$ and
 ${\displaystyle g_m(\beta)  =   \left({1+{\frac{t^{\beta}_{\alpha}}{b_{\alpha}(m,\beta)}}}\right)^{-m}}.$
Then, renaming $b({\bf \tilde{t}_m},\beta)$ in $b_{\alpha}(m,\beta)$ in (\ref{ideal.eta}-\ref{ideal.beta}),
\begin{eqnarray*}
 P_{\pi}(T\leq t_{\alpha}) & = & \iint_{\R_+\times\R_+}  F_W(t_{\alpha}|\eta,\beta) \pi(\eta|\beta)\pi(\beta) \ d\eta d\beta, \\
 & = & 1 -  {\displaystyle \Delta_{m} \left.\int_{\R_+} f_m(\beta) g_m(\beta) \ d\beta \right.} \\
\end{eqnarray*}
where $\Delta^{-1}_m = \int_{\R_+} f_m(\beta) \ d\beta$ which must
be necessarily finite to get a proper $\pi(\beta)$. Assuming $
P_{\pi}(T\leq t_{\alpha})=\alpha$ leads to
$\int_{\R_+} f_m(\beta) h_m(\beta) \ d\beta   =  0$
where $h_m(\beta)  =   g_m(\beta) - (1-\alpha)$ which is
continuous in $\beta\geq 0$ if $\beta\mapsto
b_{\alpha}(m,\beta)$ is assumed continuous for any $m>0$. Since $\beta\mapsto
f_m(\beta)>0$ except possibly on a finite number of points in
$\R_{+}$, $h_m(\beta)=0$ almost everywhere in $\R_{+}$. Expression
(\ref{choice.b}) follows immediately. \\ 

\begin{lemmo}\label{lemma1} Denote ${\cal{M}}_1$ and ${\cal{M}}_2$ two
sampling models with same parameter $\theta$. Denote ${\bf
\tilde{t}_{m_i}}$ a non-observed ${\cal{M}}_1-$sample with
likelihood ${\ell}_{{\cal{M}}^{(1)}}$. Let ${\bf
\tilde{s}_{m_i}}=g({\bf \tilde{t}_{m_i}})$ be an observed
${\cal{M}}_2-$sample with likelihood ${\ell}_{{\cal{M}}^{(2)}}$,
such that $\theta \mapsto {\ell}_{{\cal{M}}^{(2)}}(\theta) \propto
{\ell}_{{\cal{M}}^{(1)}_i}(\theta)$. Denote $\pi^J$ a prior
measure on $\theta$ and $\pi^J_{{\cal{M}}^{(j)}}$ its posterior
knowing likelihood ${\ell}_{{\cal{M}}^{(j)}}$. Then for $p$
various samples ${\bf \tilde{t}_{m_1}},\ldots,{\bf
\tilde{t}_{m_p}}$, one has
\begin{eqnarray*}
\pi^J_{{\cal{M}}^{(1)}}\left(\theta|{\bf
\tilde{t}_{m_1}},\ldots,{\bf \tilde{t}_{m_p}}\right) & = &
\pi^J_{{\cal{M}}^{(2)}}\left(\theta|{\bf
\tilde{s}_{m_1}},\ldots,{\bf
\tilde{s}_{m_p}}\right)\end{eqnarray*}
\end{lemmo}

\paragraph{Proof.}
The proof is straightforward, and  we can consider only two
non-observed samples ${\bf \tilde{t}_{m_1}}$ and ${\bf
\tilde{t}_{m_2}}$ (corresponding possibly to two independent
expert opinions). Then
\begin{eqnarray*}
\pi^J_{{\cal{M}}^{(1)}}\left(\theta|{\bf \tilde{t}_{m_1}},{\bf
\tilde{t}_{m_2}}\right) & \propto & \ell_{{\cal{M}}^{(1)}}\left(
{\bf \tilde{t}_{m_1}}|\theta\right)
\pi^J_{{\cal{M}}^{(1)}}\left(\theta|{\bf
\tilde{t}_{m_2}}\right), \\
& \propto & \ell_{{\cal{M}}^{(2)}}\left( {\bf
\tilde{s}_{m_1}}|\theta\right) \pi^J(\theta)
\ell_{{\cal{M}}^{(1)}}\left( {\bf
\tilde{t}_{m_2}}|\theta\right), \\
& \propto & \ell_{{\cal{M}}^{(2)}}\left( {\bf
\tilde{s}_{m_1}}|\theta\right) \ell_{{\cal{M}}^{(2)}}\left( {\bf
\tilde{s}_{m_2}}|\theta\right) \pi^J(\theta), \\
& \propto & \ell_{{\cal{M}}^{(2)}}\left( {\bf
\tilde{s}_{m_1}}|\theta\right)
\pi^J_{{\cal{M}}^{(2)}}\left(\theta|{\bf
\tilde{s}_{m_2}}\right), \\
& \propto & \pi^J_{{\cal{M}}^{(2)}}\left(\theta|{\bf
\tilde{s}_{m_1}},{\bf \tilde{s}_{m_2}}\right)
\end{eqnarray*}
and the equality follows by immediate normalization. 

\paragraph{Proof of Proposition \ref{prop2}.}
The Weibull likelihood of the $m$ virtual data summarized by
$(t_{\alpha},\alpha,\widetilde{\beta}(m))$ is
proportional to $\ell(\eta,\beta) =
\pi(\eta,\beta)/\pi^J(\eta,\beta)$ where $\pi(\eta,\beta)$ is
defined by (\ref{fine.eta}-\ref{fine.beta}). Modify
the parametrization using $\mu=\eta^{-\beta}$. Thus
\begin{eqnarray*}
\ell(\mu,\beta) & \propto & \mu^{m} \exp\left(-\mu
b_{\alpha}(m,\beta)\right) \beta^m
\exp\left(-m\frac{\beta}{\widetilde{\beta}(m)}\right), \\
& \propto & \left[{\left\{\mu
\left(t_{\alpha}\right)^\beta\right\}^m} \exp\left\{-\mu
k_{\alpha,m} \left(t_{\alpha}\right)^\beta\right\} \right] \
\left[ {\beta^m}
\exp\left\{-m\frac{\beta}{\widetilde{\beta}(m)}\right\}
\exp\left\{-m \beta \log t_{\alpha}\right\} \right], \\
& \propto & \ell_1\left(k_{\alpha,m}|\mu,\beta\right) \
\ell_2\left(\beta_{t_{\alpha},m}|\beta\right)
\end{eqnarray*} where
\begin{eqnarray*}
\ell_1\left(k_{\alpha,m}|\mu,\beta\right) & \propto &
\frac{\left\{\mu
\left(t_{\alpha}\right)^\beta\right\}^m}{\Gamma(m)}
k^{m-1}_{\alpha,m} \exp\left\{-k_{\alpha,m}
\left(t_{\alpha}\right)^\beta\right\}
\end{eqnarray*}
which is the likelihood arising from considering
$k_{\alpha,m}\sim{\cal{G}}(m,m\mu (t_{\alpha})^\beta)$, and
\begin{eqnarray*}
\ell_2\left(\beta_{t_{\alpha},m}|\beta\right) & \propto &
\frac{\left(m\beta\right)^m}{\Gamma(m)} \beta^{m-1}_{t_{\alpha},m}
\exp\left\{-m\beta\frac{1+\widetilde{\beta}(m)\log
t_{\alpha}}{\widetilde{\beta}(m)}\right\}
\end{eqnarray*}
which is the likelihood arising from considering
$\beta_{t_{\alpha},\delta}\sim {\cal{IG}}(m,m\beta)$. \\

\end{document}